# Presenting a new approach in security in inter-vehicle networks (VANET)

1st Davoud Yousefi, 2nd Farhang Farhad, 3st Mehran Abed, 4nd Soheil Gavidel

1. Department of Computer Engineering, Moghadas Ardabili Institute of Higher Education, Ardabil, Iran
   Email: d.yousefi.sh@gmail.com
2. Department of Computer Engineering, Moghadas Ardabili Institute of Higher Education, Ardabil, Iran
3. Department of Computer Engineering, Moghadas Ardabili Institute of Higher Education, Ardabil, Iran
4. Department of Computer Engineering, Moghadas Ardabili Institute of Higher Education, Ardabil, Iran

**Abstract**

Nowadays, inter-vehicle networks are a viable communication scenario that greatly contributes to daily work, and its issues are gaining more and more attention every day. These days, space networks are growing and developing. There are numerous new uses for this new kind of network communication. One of the most significant daily programs in the world today is road traffic. For human growth, passenger and freight transportation is essential. Thus, fresh advancements in the areas of improved safety features, environmentally friendly fuel, etc., are developed daily. In order to improve safety and regulate traffic, a new application program is used. However, because of their stringent security standards, these initiatives have an impact on traffic safety.

Since driving is one of the things that necessitates traffic safety, this area needs to be made more secure. Providing trustworthy driving data is crucial to achieving this goal, aside from the automated portion of the operation. Drivers would greatly benefit from accurate weather descriptions or early warnings of potential dangers (such as traffic bottlenecks or accidents). Inter-vehicle networks, a novel form of information technology, are being developed for this reason.

## I. Introduction

Road safety, life safety, traffic control, and passenger and driver enjoyment can all be facilitated by vehicle communication networks, an extremely promising and new technology. The inter-vehicular network is an example of a smart transportation system that uses roadside infrastructure to facilitate communication between cars. Among the many services and advantages this network offers its customers are ease of access to the Internet, navigation, toll payment services, and a reduction in traffic on the highways, all of which can improve the driving experience for drivers. Other uses are also available, such as diversion information, road conditions (like slippage), emergency vehicle alarm signals (like ambulances), and different warning messages to prevent traffic [1].

The 75MHz bandwidth is allocated in the 925/5-850/5 GHz band, and vehicles use a proprietary short-range communications protocol for communication. The range of communications using short-range communications protocols is over ۵۰۰ meters for vehicles and 1000 meters for roadside infrastructure, and a vehicle can transmit each message in 100-300milliseconds. Rusted. The space network system has a large number of nodes, more than 750million in the world. Each node can move freely in any direction and remain connected to the network. The communication between nodes can be direct or indirect and through one or more jumps between nodes [2].

If such a network is not secure before deployment, it is at high risk. For example, if a safety message is changed, discarded, or delayed intentionally or due to a hardware problem, serious consequences such as injury and even death may occur. This requires and demands the development of a functional, reliable, efficient and security architecture before implementing all aspects of the inter-vehicle network [5]. The establishment of security in the inter-vehicle network pursues specific objectives, including 1. ensuring the correctness of the information and data received (information authentication), 2. source authentication 3. information confidentiality Personal node sender and 4. resistance to attacks pointed [3].

## II. VENT NETWORK

A form of network (WANT) is a vehicle case network that provides communication between wireless case MANETs of vehicles in close proximity, between vehicles and fixed equipment along the route, known as RSU, equipment Each node, for example, a vehicle or roadside equipment, is connected to other nodes in one or more stages. Passenger VANET is designed as follows [4]:

A. Vehicle-to-vehicle communication or (V-V).

B. Communications between vehicles. Vehicle communication with infrastructure or (V-I).

C. Vehicle communication with roadside equipment4) Communication between roadside equipment.

Vehicle-to-vehicle case networks (VANETs) are examples of mobile case networks, in which each vehicle acts as a smart node in which vehicles with infrastructure equipped with transmission capabilities form A network is connected [5]. Vent's research is aimed at revolutionizing the delivery of multi-step messages from a moving vehicle to the network infrastructure. One of the inherent characteristics of the van is the high mobility of vehicles, which leads to topology changes and instability of links during short communication times between vehicles. The opportunistic network is one of the topics that has received a lot of attention from the scientific community in the last decade. In these networks, mobile nodes are able to communicate asynchronously and there is often no end-to-end path between the origin and the destination. Routing in the van has the characteristics of energy constraints and high speed of vehicles, which is one of the challenging issues in the opportunistic routing algorithm. In this algorithm, data is sent to vehicles based on the carrying storage mechanism and then sent. Most opportunistic routing algorithms use message flooding to increase delivery rates, while ignoring the overhead caused by it, but one of these routing methods can meet this need and it is the application of the single-cast strategy to solve the problem, which decides the next step based on the route of the carrier vehicle and the target vehicle [6].

## III. TYPES OF INTER-VEHICLE NETWORKS

Intelligent inter-vehicle networks are very effective in creating an intelligent transportation system, which are an example of the use of artificial intelligence in cars and cause the appearance of intelligent autonomous behavior by cars in situations such as accidents, driving while drunk, become drivers and so on. If we want to divide car networks based on information transmission media, we reach three general categories: networks based on radio waves, based on positioning and based on sensors. VANET networks based on radio waves use different wireless technologies such as DSRC, which is a type of Wi-Fi. It also uses cellular and WiMAX technology. Other short-band wireless protocols including IEEE 802.11 Blue tooth and CALM can be used in this type of networks. The use of different routing protocols such as the GPSR routing protocol is one of the leading protocols in location-based networks, and DSR and DSDV are being investigated in VANET networks, but it seems that the routing protocols of ad hoc AODV and DYMO are more efficient in this type of networks [7].

The sensors, which must be employed in various areas of the vehicle and send the vehicle's condition and outside surroundings to the driver and controller, are the most crucial component of the VANET network. The controller can then use the driver's directives or data from other vehicles [8].

Sensors in VANET are divided into two types: autonomous sensors and cooperative sensors. Autonomous sensors include ART sensors to measure the radio range and vehicle communication with other network nodes, MGT to measure the maximum speed that the driver can have and still remain in the network, MDT to detect vehicles that are physically present in the network, and sensors The other is to identify the position of the car on the map and determine the distance with the front and rear cars. Participating sensors are used to announce changes in the status of neighboring vehicles, reach a specific location, receive information needed by the driver, etc. [8].

### A. Vehicle network architecture

With the increase of vehicles on the roads, the appropriate speed and safety distance of passengers has decreased and driving has become more dangerous. Vehicle communication is V2V and V2I. V2V is a vehicle-to-vehicle communication, where one vehicle broadcasts information to other vehicles. In V2I, these connections are with roadside equipment, cars can pay highway fees without stopping [9].

### B. On-board unit (OBU)

In the car network, mobile equipment called OBU is installed in the cars so that it is possible to communicate with the network, and at certain times, messages related to traffic, speed, brake status, road signs and direction of movement are sent to other cars. broadcast and messages related to other vehicles are sent and processed to the desired vehicle, there is a WAVE device to exchange information with OBUs and roadside units. Usually, OBU is connected to a memory user interface, processor, a link interface with OBUs, or connected to RUS through wireless networks. OBU tasks are geographic routing, message transmission, secure radio access, information security, IP mobility, and congestion control [10].

### C. Application unit (AU)

AUs are equipped devices inside the car that use services. They do this by using OBU capabilities [11].

## D. Roadside units (RSU)

RSUs are WAVE devices that are fixed in places such as parking lots or in different parts of the road. RSUs are targeted for communication within the network infrastructure and are equipped with devices dedicated for short-range communication based on radio technology. According to the consortium, the main operations related to RSU are: expanding the range of network communication, by redistributing messages from OBUs and amplifying the message to RSUs and transferring it to OBUs and implements applications for safety purposes [12–14].

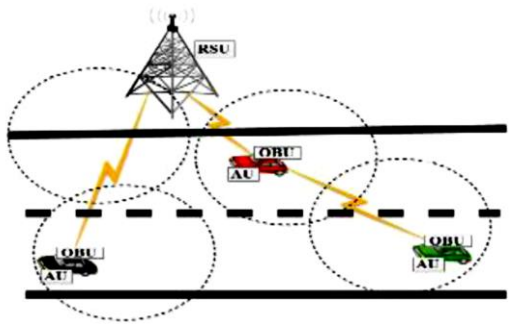

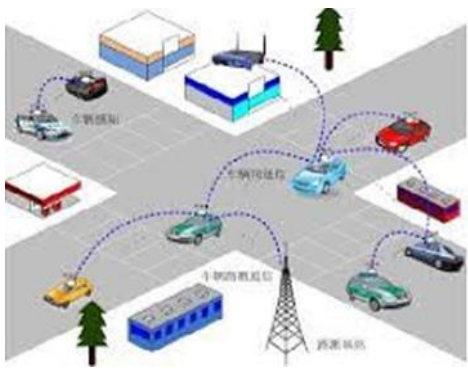

Figure 2.1. The scope of communication of automotive networks is expanded to other OBUs through the redistribution of message [15].

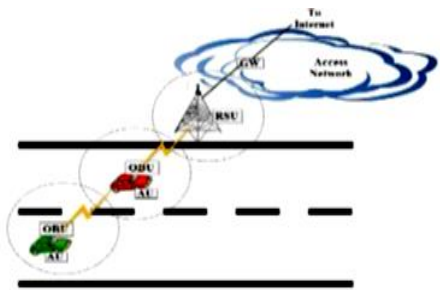

Figure 2.2. RSU provides OBUs with Internet connections [16].

## E. Network challenges of inter-vehicle networks

The most important challenge of inter-vehicle networks is the matter of security. It is easy to imagine that if the smallest threat is directed at such networks, there will be life and financial consequences, and unlike the Internet or the web, human health and lives are at stake. is directly exposed to danger. It is enough that the inter-vehicle network does not provide accurate information to the driver and this causes accidents and overturns and other risks. Networks of inter-vehicle networks are the biggest transforming technology of cars that gives security to drivers [17].

Various solutions based on information encryption between cars, isolating the network and using authentication mechanisms for new cars that want to enter the network, designing dedicated validation programs such as ELP and VPKI and using reliable hardware 99.99 percent and dozens of other plans are some of the works done to secure the networks of inter-vehicle networks. But apart from security, the next important challenge is routing information and communication between cars at high speed. Due to the lack of a central node, a car cannot directly send and receive information and must use intermediate cars to communicate with another car, and because speed is a very critical factor in these networks, the routing algorithm is of double importance [18].

### Findings

With the introduction of the knowledge of communication technology in the world of automobile manufacturing, automobile manufacturing companies put their best efforts in building communication equipment inside the car to increase safety, reduce resource wastage, correct routing to reduce time wastage and improve the driving experience and prevent Traffic, as well as the use of entertainment facilities and comfort, have increased the number of people. The working mechanism of inter-vehicle networks is such that when braking, the communication system of inter-vehicle networks sends a warning message and automatically reduces the speed of the vehicle. Of course, in inter-vehicle networks, various messages are periodically published in order to create the necessary space for safe driving.

### A. Applications and requirements of inter-vehicle networks

Why are inter-vehicle networks so important? What are the advantages and disadvantages compared to cellular and dedicated networks that have caused them to be developed separately and car manufacturing companies to invest heavily in them and countries such as the United States to approve the necessary standards, protocols and licenses. and define and use them in intelligent transportation systems of advanced countries? Inter-vehicle networks are generally used for three primary and distinct purposes: business, convenience, and safety. Case networks' built-in features, like short-range optimal application for traffic problems and vehicle safety is created by high-speed network formation, topology changes, and signal transmission from the source to the destination. In just a few hundred meters, cars can be alerted to an accident and alter their course. They can also detect traffic and ask questions of the car in front of them or behind them, as well as about the traffic conditions on the street or side alleyways, and they can learn about the next crossroads. Driving may be safer, more comfortable, and more pleasurable when drivers use the information, they receive from the cars around them to make better judgments [19].

In inclement weather, the uses of inter-vehicle networks become more obvious. Traffic control center alerts are sent by RHCN, accident warnings are sent by PCN, collision warnings are announced by CCW, and the security software SVA is utilized to warn, slow down, or stop the vehicle in inter-vehicle networks. EEBL is used for sudden braking. The traffic department uses TOLL to pay tolls without stopping the vehicle, PAN for parking warnings, and CRN programs for traffic warnings. Inter-vehicle networks can serve intriguing purposes from a business perspective.

The vehicle can be customized via the RVP/D program to suit the driver's requirements and preferences. Through the main roadside stations, the driver or vehicle can access their favorite music, public city information or radio channels, games, and other forms of amusement. A different application named SA has the ability to broadcast various announcements in the automobile based on the demands of the driver [20].

### B. Types of communication in inter-vehicle networks

Vehicle-to-vehicle (V2V), Vehicle-to-Roadside (VRV), and Vehicle-to-Infrastructure (V2I) communication will become more and more imperative as networks and mobile wireless devices gain prominence. In relation to vehicles, a safe driving environment is provided and away from traffic jams, and they provide accident information, weather information and hazards and reduce travel time. They exchange important warning and safety messages that can be reported before any accident and road traffic congestion. This process includes vehicles (mobile nodes), RSUs (roadside units), TAs (trusted authorities). Vehicle communication has the ability to exchange local information in real time to enhance safe driving and improve the movement of nodes. There are many characteristics of inter-vehicle networks such as short communication, significant computing power, high mobility, dynamic topology, etc. Due to the topology of continuous change between vehicles, the safety of routing protocols is considered as a great challenge task for researchers. In inter-vehicle networks, there is also a malicious node that can change and spread fake messages in the network to cause accidents. In this regard, a reliable mechanism between nodes is needed to avoid the sharing of fake messages in inter-vehicle networks [21].

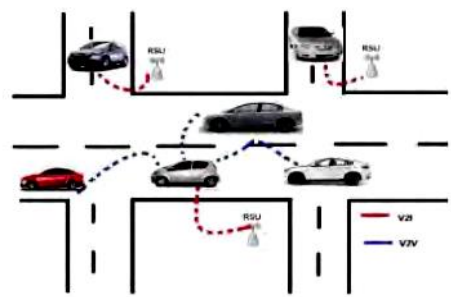

Figure 2.3. A view of the types of communication in inter-vehicle networks [22]

Many reliable and secure routing protocols have been implemented to ensure message integrity and calculate the confidence value of each message content. The trust model creates a relationship between all the neighboring nodes and their trust level recommendation. In addition, it efficiently identifies malicious nodes and solves security issues related to node failure [22].

### C. Advantages and disadvantages of inter-vehicle networks

The goal of inter-vehicular network systems is to provide a platform for various applications that can improve traffic safety and productivity, driver access, traffic regulation, entertainment, etc. There are significant research and industrial efforts for development in the field of this market. The main driving force behind inter-vehicle networks. It is for the purpose of improving security and traffic efficiency. Inter-vehicle networks allow the vehicle to automatically notify the nearby vehicle about actions such as stopping, changing lanes, etc., in order to prevent dangerous situations. Although these warning messages only require a broadcast range of less than 100 meters, they also have a very strict requirement to be processed quickly in real-time. In order for other cars to utilize the route information, inter-vehicle networks also enable vehicles to notify other vehicles about road conditions, such as traffic and accidents. No difficult points to decide on. There are some notifications that need a greater broadcast range [23].

However, the emphasis on their need for real-time processing is much less than that of alerts. This relaxation of time constraints and computing power of OBUs allows creating secure and reliable notification messages using advanced encryption. While great benefits are expected from space communication and the large number of media are the strengths of inter-vehicle networks, there are still problems in establishing such networks in practice. One of its most important problems is related to guaranteeing the security of the notification generating media. In the case of security, a selfish vehicle may try to clear the road ahead or confuse the road behind with false traffic announcements, criminals in pursuit may make fake announcements to the media to block the way of the police. publish others. Such attacks may cause serious injuries or even death. Another problem is privacy protection. is the media. Inter-vehicle networks open a big window for observers. It is very easy to obtain information about the speed, status, route and storage location of vehicles in inter-vehicle networks. By exploring this information, malicious observers can draw conclusions about the driver's personality (for example, someone who drives slowly is likely to be a quiet person), life, habits and social relationships (the places people go to). gives a lot of information about their lives). This private information may be traded in underground markets, exposing vehicles and drivers to harassment (e.g. unwanted advertisements, threats, e.g. when the driver often goes to an embarrassing place such as a red light district) and dangers. ) like theft) to be placed. Finally, inter-vehicle networks are especially attractive in densely populated urban areas and drowned in traffic and accidents. Apart from this vulnerability to attacks against traffic security and privacy of drivers, large-scale inter-vehicle networks in urban areas increase the scalability and ability to manage problems [24].

### D. An overview of security in inter-vehicle networks

Inter-Vehicle Networks (VNET) are a new form of ad hoc networks, the construction of such a protocol leads to faster transmission of data and messages in vehicles and roadside vehicles. Messages and information are transferred from one car to other cars in the inter-car network using wireless devices, which increases safety and optimizes traffic on the roads. Therefore, the main challenge in the inter-vehicle network is the effective data path from the origin to the destination. Due to the easy accessibility of this network, the possibility of problems by malicious users is high, therefore, security is the first word in this network. The discussion about security issues in space networks is the main topic of this article. In this chapter, we will introduce the challenges of this area and a set of solutions to solve problems in this network [25].

### E. Inter-vehicle network security

The security of inter-vehicle network is very important and the main goal of security in inter-vehicle network is to protect information and resources against attacks that a security system for the security of messages must be able to neutralize any public attack on the network. Despite the recent developments in the inter-vehicle network, many channels can attack and endanger this network because of their power beyond the inter-vehicle network. Raya et al. classified attackers in three categories: insider vs. outsider, destructive vs. logical, and active vs. passive. There are different types of attacks in inter-vehicle networks, which we will mention here [26-27].

### F. The importance of providing security in inter-vehicle networks

The main goal of inter-vehicle systems is to create security and comfort for drivers and passengers. The issue of security in VANET networks is so sensitive and critical that it is one of the reasons for the non-implementation of these networks in the past years. to be A variety of solutions based on encryption of information between cars, isolating the network and using authentication mechanisms for new cars that want to enter the network, designing special validation programs such as ELP and VPKI and using reliable hardware. 99.99% and dozens of other schemes are some of the work done to secure inter-vehicle networks. Critical information cannot be changed or deleted by an attacker. ¬Gives the ability to determine the driver's responsibility while preserving the driver's privacy. To exchange information about vehicles and drivers in a safe and timely manner, the delay in message exchange may cause catastrophic consequences such as vehicle collisions [28].

### G. Security requirements

Concerns over the accuracy of a system's information are raised when data is gathered and disseminated among anonymous users.

- In order for drivers to be unable to dispute that their vehicles are the source of these messages and to give inspectors the opportunity to ascertain legal eligibility, it should be feasible to reliably identify the vehicles that pass through the automated core messages. An accident occurs [29–34].

- Exclusive: the privacy of drivers in front of unauthorized observers should be guaranteed, in this case the abuse of automatic nuclear messages to track the movements of cars is prevented.

- Access: The space network must be available to everyone at all times. To use this program, there may be an urgent need for this program and a faster response than the network or even ad hoc, so a delay even for a second may be meaningless. And even destructive.

- Privacy: It is an important principle to keep the driver's information from unauthorized observers, such as real identity, travel route, speed, etc.

- Real time limitations: Vehicles move at high speed, a real and limited time is needed to respond in some situations and the lack of this limit will have devastating results.

- Validation: Since cars must only react to legitimate messages from the auto core, these messages must be validated.

Data compatibility: In order to protect against wrong automatic kernel messages that are generated from legitimate machines (technical equipment or software defects), it must be possible to verify the compatibility and compatibility of kernel messages. A reliable communication should generally have two characteristics: 1) the sender is finally accepted as a reliable source; 2) Sender's message should not be tampered with during transmission. An important requirement in VANET security is the validation of the sender in communication [31].

### H. Physical security challenges

VANET's have been exposed to various threats and attacks. Since the vehicle itself does not have a sufficient source of electricity, the OBU and the battery life is limited, they use mobile devices such as smart phones and wearable devices. It is not possible to send and receive directly, it has to use intermediate vehicles and because the speed in these networks is a very important factor, the routing algorithm becomes doubly important. Among the multiple signals, the mechanisms for detecting the presence and disappearance of cars in a network support the standards and protocols of automobile manufacturers and the communication between sensors are other challenges of VANET [32].

### I. types of attacks

Denial of service attacks: This attack occurs when the attacker takes time to control the resources of the communication devices or the communication channel used by the space network, after which it prevents the entry of critical information because the driver relies on the information of the program. This increases the risk, for example, if a malicious person wants to create a huge problem on the highways, he can create an accident and use DOS attacks to prevent vehicles from approaching [35].

### J. Credit threat

Credibility in a space network includes the protection of legitimate nodes from the penetration of internal or external attacks into the network using fake identity, identifying suppressed attacks, creating, changing or broadcasting legitimate messages, revealing fake global positioning system signals and preventing the introduction of false information to the space network. Here we mention a number of credit threats [36].

### K. The threat of reliability

By using techniques like illegal message collection through eavesdropping and broadcast message collection to gather situational information, the private communication between the nodes of a space network is at risk. When it comes to eavesdropping, both internal and external attackers have the ability to gather data about road users and utilize it without the user's knowledge. Anonymity and privacy are significant concerns for drivers. The safety of users by concealing their precise position in space and time is part of maintaining location privacy. A degree of anonymity has been attained by concealing a user's request in a way that makes it unidentifiable to other users [37].

### L. Identity verification with digital signature

Digital signature authentication is a smart option for inter-vehicle networks. Additionally, public key infrastructure—where each vehicle is given a pair of private and public keys—is an excellent technique of implementing authentication due to the huge number of network participants and changeable connectivity to authentication servers [38–39].

### Conclusion and suggestions

Today's daily job greatly benefits from inter-vehicle networks, a potential communication scenario whose issues are gaining more and more attention every day. Space networks are growing and changing today. This new kind of network communication offers a number of new uses. These days, one of the most significant everyday programs worldwide is road traffic activities. Transportation of people and goods is essential to human progress. As a result, improvements in safety features, environmentally friendly fuel, etc., are being made daily. As a result, a new application software is used, improving safety and traffic management. Road traffic safety is impacted by these initiatives, nevertheless, because of their stringent security requirements.

Driving is one of the activities that require traffic safety, so this area needs to be made more secure. In addition to the automated part of the operation, providing reliable driving data is essential to accomplishing this goal, as drivers would greatly benefit from accurate weather descriptions or early warnings of potential dangers (e.g., traffic bottlenecks or accidents). For this reason, a new type of information technology called inter-vehicle networks is being developed.

The inter-vehicle network demonstrates that some of the special characteristics of conventional security mechanisms—such as high node mobility, geographic growth, etc.—are not necessarily appropriate. As a result, a number of common studies have been offered thus far. In addition, inter-vehicle networks are facing many security concerns that are continually becoming more prominent. There are a number of potential future studies in the developing field of inter-vehicle network security. Some concerns need be addressed despite the fact that a number of approaches have been offered (for example, privacy difficulties with respect to radio frequency fingerprinting).

In addition, different vehicular network protocols, mechanisms, and applications depend on different architectures and assumptions, so a common evaluation framework is needed to compare the role of different security research. In this seminar, the types of risks that have arisen in the various security areas of this network have been investigated and some ways to deal with these attacks have been briefly presented. In turn, it has become a network platform to support future applications. Safety and security are becoming a necessity for vehicular ad hoc network applications. The vehicular ad hoc network as one of the networks that uses wireless technologies has made it vulnerable to many risks. In this article, we first discussed the architecture and applications of this case network and then discussed the importance of providing security, security goals and finally the types of attacks in the case network of vehicles.